\documentclass[%
 reprint,
 amsmath,amssymb,
 aip,jcp,
 nofootinbib,
]{revtex4-1}

\usepackage[T1]{fontenc}
\usepackage[utf8]{inputenc}
\usepackage{graphicx}
\usepackage{hyperref}
\usepackage[capitalize]{cleveref}
\usepackage[english]{babel}
\usepackage{todonotes}
\usepackage{siunitx}
\usepackage{dsfont}
\usepackage{bm}

\date{\today}
\graphicspath{{./}{figures/}{plots/}}

\newcommand{\eqdot}{.}
\newcommand{\eqcomma}{,}

\newcommand{\agrid}{a_\text{grid}}
\newcommand{\tgrid}{\tau}
\let\vec\mathbf
\newcommand{\tens}[1]{\bar{\bm{#1}}}

\renewcommand\citep[1]{%
 {%
   \let\textsuperscript\relax
   [\cite{#1}]%
 }%
}

\begin{document}
\title{Moving Charged Particles in Lattice Boltzmann-Based Electrokinetics}
\author{Michael Kuron}
\email{mkuron@icp.uni-stuttgart.de}
\author{Georg Rempfer}
\affiliation{Institut f\"ur Computerphysik, Universit\"at Stuttgart, 70550 Stuttgart, Germany}
\author{Florian Schornbaum}
\author{Martin Bauer}
\author{Christian Godenschwager}
\affiliation{Lehrstuhl f\"ur Systemsimulation, Friedrich-Alexander-Universit\"at Erlangen-N\"urnberg, 91058 Erlangen, Germany}
\author{Christian Holm}
\affiliation{Institut f\"ur Computerphysik, Universit\"at Stuttgart, 70550 Stuttgart, Germany}
\author{Joost de Graaf}
\email{j.degraaf@ed.ac.uk}
\affiliation{Institut f\"ur Computerphysik, Universit\"at Stuttgart, 70550 Stuttgart, Germany}
\affiliation{Current address: SUPA, School of Physics and Astronomy, The University of Edinburgh, Edinburgh, EH9 3FD, United Kingdom}

\begin{abstract}
The motion of ionic solutes and charged particles under the influence of an electric field and the ensuing hydrodynamic flow of the underlying solvent is ubiquitous in aqueous colloidal suspensions.
The physics of such systems is described by a coupled set of differential equations, along with boundary conditions, collectively referred to as the electrokinetic equations.
Capuani~\textit{et al.} [J.~Chem. Phys. \textbf{121}, 973 (2004)] introduced a lattice-based method for solving this system of equations, which builds upon the lattice Boltzmann algorithm for the simulation of hydrodynamic flow and exploits computational locality.
However, thus far, a description of how to incorporate moving boundary conditions into the Capuani scheme has been lacking. Moving boundary conditions are needed to simulate multiple arbitrarily-moving colloids.
In this paper, we detail how to introduce such a particle coupling scheme, based on an analogue to the moving boundary method for the pure LB solver.
The key ingredients in our method are mass and charge conservation for the solute species and a partial-volume smoothing of the solute fluxes to minimize discretization artifacts.
We demonstrate our algorithm's effectiveness by simulating the electrophoresis of charged spheres in an external field; for a single sphere we compare to the equivalent electro-osmotic (co-moving) problem.
Our method's efficiency and ease of implementation should prove beneficial to future simulations of the dynamics in a wide range of complex nanoscopic and colloidal systems that was previously inaccessible to lattice-based continuum algorithms.
\end{abstract}

\maketitle

\section{Introduction}

The dynamics of electrolytic solutions is essential in the description of most most processes in (bio)chemistry and soft matter physics.
This includes electrostatic screening via the formation of double layers, electro-osmotic flow\cite{anderson89a,paxton06a,ibele07a,reinmuller12a}, electrophoresis\cite{obrien78a,smoluchowski17a,hueckel24a,henry31a}, and self-electrophoresis\cite{paxton04a,moran11a,moran14a,brown14a,ebbens14a,sabass12b,brown15a}.
Industrial applications range from DNA sequencing\cite{woolley95a,mikheyev14a} and oil recovery\cite{pagonabarraga10a} to the detection\cite{castellanos01a}, separation\cite{schaegger87a}, and characterization\cite{blanche81a} of analyte molecules.
The coupled occurrence of diffusion, hydrodynamics, and electrostatics, often referred to as electrokinetics or electrohydrodynamics, gives rise to complex physical behavior.
Electrokinetic processes involve dynamics on vastly different length scales with double layers typically having a nanometer scale in aqueous solution and system geometries often ranging from several micrometers to millimeters.
The small length scales make it difficult to access the nanoscopic details in experiment, while the large discrepancy in length scales poses a challenge to simulation methods.

Existing methods that solve the electrokinetic equations numerically are based on a variety of algorithms for hydrodynamics, generally falling into two classes.
Firstly, particle-based algorithms, which include dissipative particle dynamics\cite{medina_efficient_2015,smiatek11a} (DPD) and multi-particle collision dynamics\cite{frank08a,frank09a} (MPCD).
Secondly, continuum (grid-based) algorithms, which encompass methods like finite elements\cite{moran11a,moran14a,brown15a,lewis72a} (FEM), the boundary element method\cite{allison05a,hoyt98a,house12a} (BEM), finite differences\cite{schmitz12a}, finite volumes\cite{jeong03a}, electrokinetics based on lattice Boltzmann (LB)\cite{capuani04a,warren97a}, and the smooth profile method \cite{nakayama08a} (SP).
For additional details, we refer to the overviews given in Refs.~\onlinecite{pagonabarraga10a,rotenberg13a}.

The particle-based methods solve the full time-dependent problem and are intrinsically able to include mobile charged colloids and macromolecules.
The downside of these methods is that solute ions, water, and macromolecules are resolved on the same scale.
This sets a low limit for the maximum size of the simulation domain, given current computational abilities.

Continuum solvers allow for the study of systems with much larger length scales because the ions are not resolved.
In general, introducing moving boundaries in numeric solvers for the continuum electrokinetic equations poses a more challenging problem than for their particle-based counterparts.
Tackling this issue, however, is well worth the effort, as it grants access to the wide variety of physical problems
beyond those that can be studied via Galilean transformation to the rest frame of the charged object, an approach that is not possible in general.
Moving particles will permit the study of interactions of multiple (self-)electrophoretically driven colloids, as well as their interactions with any stationary boundaries in the system.

When it comes to resolving large system sizes, the FEM method has proven itself to be a very efficient solver for the stationary electrokinetic equations.
FEM has been successfully applied in arbitrary geometries and for a diverse range of physical systems, including: nanopores\cite{white_ion_2008,lan_nanoparticle_2011,rempfer16b}, self-electrophoresing colloids\cite{moran11a,moran14a,kreissl16a,brown15a,qian08b,wang13b,chiang14a}, and electro-osmotic pumping\cite{mengeaud02a,bianchi00a}.
The strength of the FEM lies in its ability to adaptively increase the resolution in regions with high gradients, i.e., the electric double layers.
The downside is that the treatment of time-dependent boundary positions is computationally exceptionally expensive, due to the constant need for remeshing.
However, there are examples of its application to phoretic problems with moving boundaries\cite{yang16a}.

An alternative to FEM was proposed by \citet{capuani04a}. Their lattice electrokinetics method (EK) combines finite differences and finite volumes to solve the diffusion-advection of solute and the electrostatics problem.
For hydrodynamics, they exploit the computational efficiency of the LB method\cite{mcnamara88a,duenweg09a}.
Unlike earlier methods that aimed to achieve a similar level of description\cite{warren97a}, EK propagates the solute by considering solute fluxes between neighboring lattice cells instead of concentrations in lattice cells, thus ensuring local mass conservation and reducing discretization artifacts\cite{capuani04a}.
Furthermore, EK, like LB, is a local algorithm, with the exception of the electrostatics solver, making it generally more efficient than FEM and other solvers, which are fully nonlocal.

Of specific interest is EK's ability to resolve the particles' surfaces as boundary conditions on the grid.
This allows for the study of extended objects and straightforward incorporation of local variation of the surface properties, which is, e.g., useful in studying self-propelled particles \cite{scagliarini16a}.
Extending this method further with dynamically-adaptive grid refinement, a common practice in LB\cite{schornbaum16a}, will constitute an algorithm fully capable of directly modeling microfluidic systems.

The EK method has previously been used to study problems such as electrophoresis\cite{giupponi11a} and sedimentation\cite{capuani04a} of single charged spherical colloids.
Other applications include charge transport in porous media\cite{rotenberg08a,obliger14a} and the translocation of DNA molecules through nanopores\cite{reboux06a}, where this method can tremendously reduce computational effort when compared to particle-based methods, and thus enable one to reach experimentally relevant length and time scales.


In this manuscript, we present an extension of the lattice electrokinetics method proposed by \citet{capuani04a} to include moving boundaries.
For the underlying LB method, a moving boundary scheme\cite{ladd94a,aidun98a} is already widely employed for moving particles.
To adapt the EK method for moving boundaries, we first introduce solute mass and charge conservation by displacing solute from cells that are claimed by a moving particle.
We subsequently add a partial volume smoothing to reduce the effect of a cell being claimed in a single time step.
We do so by incorporating a term in the solute flux calculation that depends on the amount of volume in a cell that is overlapped by a particle.
This allows for a gradual expulsion of solute from a cell as it is increasingly overlapped.

We validate our method by considering the problem of electrophoresis of spherical colloids in an external field.
We observe that the moving boundary method is capable of reproducing results using the Capuani method with fixed boundaries (in the co-moving frame).
Moreover, we observe that the smoothing term is necessary to prevent strong variations in the particle velocity due to the grid discretization --- it reduces the oscillations by more than a factor of $10$.
The added computational cost due to the moving boundaries amounts to approximately a $10\%$ increase per time step, compared to the stationary boundary algorithm of \citet{capuani04a}.
The method we describe retains the flexibility and efficiency of the LB and EK methods, while granting the EK method access to an entire class of problems where the relative motion of multiple particles is of relevance.

The remainder of this paper is laid out as follows:
In \cref{sec:equations}, we give the governing equations.
In \cref{sec:numerical}, we introduce the numerical methods, including a novel method for coupling moving particles with EK.
In \cref{sec:validation}, we validate the new method by comparing equivalent electrophoretic and electro-osmotic systems.
We conclude in \cref{sec:conclusion}.

\section{The Electrokinetic Equations}
\label{sec:equations}

In this section, we introduce the electrokinetic equations for bulk electrolytes.
These describe the motion of a charged solute in a solvent fluid by a diffusion-advection process.
Boundary conditions will be discussed in \cref{sec:numerical}.
For each solute species $k$, the solute flux is given by
\begin{align}
	\vec{j}_k(\vec{r},t)=&
	\underbrace{-D_k\vec{\nabla}\rho_k(\vec{r},t)
	-\frac{D_k}{k_\text{B} T} z_ke \rho_k(\vec{r},t) \vec{\nabla} \Phi(\vec{r},t)}_{\vec{j}^\text{diff}_k} \nonumber\\
	&+ \underbrace{\rho_k(\vec{r},t) \vec{u}(\vec{r},t)}_{\vec{j}^\text{adv}_k}
	\label{eq:flux}
\end{align}
and obeys the continuity equation
\begin{equation}
	\frac{\partial \rho_k}{\partial t}(\vec{r},t) = - \vec{\nabla} \cdot \vec{j}_k(\vec{r},t)
	\label{eq:continuity}
	\eqdot
\end{equation}
Here, $\rho_k(\vec{r},t)$ is the concentration field and $\vec{j}_k(\vec{r},t)$ is the associated flux.
$\Phi(\vec{r},t)$ refers to the electrostatic potential and $\vec{u}(\vec{r},t)$ to the fluid velocity.
$\vec{\nabla}\cdot$ is the divergence operator; $\partial/\partial t$ denotes the time derivative.
$D_k$ is the diffusion coefficient of the $k$th solute, $k_\text{B}$ Boltzmann's constant, $T$ the absolute temperature, $z_k$ is the $k$th solute's ionic valency, and $e$ is the unit charge.

The electrostatic potential is given by Poisson's equation
\begin{align}
	\nabla^2 \Phi(\vec{r},t) &= -\frac{1}{\varepsilon} \sum\limits_k z_ke\rho_k(\vec{r},t) \nonumber \\ &= -\frac{4\pi \lambda_\text{B} k_\text{B}T}{e} \sum\limits_k z_k\rho_k(\vec{r},t)
	\eqcomma
	\label{eq:poisson}
\end{align}
with the Bjerrum length
\begin{equation}
	\lambda_\text{B}=\frac{e^2}{4\pi \varepsilon k_\text{B} T}
	\eqdot
\end{equation}
A spatially homogeneous dielectric permittivity $\varepsilon$ is assumed here.

As the relevant length scales in colloidal systems are on the nano- and micrometer scale, we are in the low Reynolds number regime --- viscous friction dominates over inertial forces --- and the dynamics of the fluid is thus described by the stationary incompressible Stokes equations: 
\begin{align}
	\eta \nabla^2\vec{u}(\vec{r},t) &= \vec{\nabla} p(\vec{r},t) - \vec{f}_\text{ext}(\vec{r},t) \eqcomma \label{eq:stokes} \\
	\vec{\nabla}\cdot\vec{u}(\vec{r},t) &= 0 \label{eq:continuity-fluid}
	\eqcomma
\end{align}
with the kinematic viscosity $\eta$, the pressure density $p(\vec{r},t)$, and the applied force density $\vec{f}_\text{ext}(\vec{r},t)$.
The fluid's continuity \cref{eq:continuity-fluid} neglects the contribution of the solute to the total mass, which is reasonable at typical solute concentrations up to several mol/l \citep{degraaf15c}.
The force density couples the motion of the fluid to the dynamics of the solutes via the following equation\cite{rempfer16a}:
\begin{equation}
	\vec{f}_\text{ext}(\vec{r},t)=k_BT\sum\limits_k \frac{\vec{j}_k^\text{diff}(\vec{r},t)}{D_k}
	\eqdot
	\label{eq:fluidcoupling}
\end{equation}
This represents a frictional coupling proportional to the relative velocities of fluid and ions.

\section{Numerical Methods}
\label{sec:numerical}

In this section, we introduce the numerical methods employed to solve the electrokinetic equations~\eqref{eq:flux}-\eqref{eq:fluidcoupling}.
\Cref{sec:lb,sec:ek} cover the LB method and the EK method, respectively, while \cref{sec:ladd} describes the moving boundary method for coupling particles into LB.
\Cref{sec:ek-moving} describes a new method for combining moving boundaries with EK.

\subsection{Lattice Boltzmann}
\label{sec:lb}

The LB method is based on solving the Boltzmann transport equation,
\begin{align}
	\frac{\mathrm{d}f}{\mathrm{d}t}=
	&\frac{\partial f}{\partial t} + \vec{v} \cdot \vec{\nabla}\!_x f + \frac{1}{m} \vec{F} \cdot \vec{\nabla}\!_v f 
	=
	\left. \frac{\partial f}{\partial t} \right| _\text{collision}
	\equiv \tens{\Omega} f
	\label{eq:transport}
	\eqdot
\end{align}
$\vec{\nabla}\!_a$ is the gradient operator with respect to $\vec{a}$.
$f=f(\vec{x},\vec{v}, t)$ is a system's single-particle phase space probability distribution function.
It gives the probability density of finding a particle at position $\vec{x}$ with a velocity $\vec{v}$ at time $t$.
$\vec{F}$ is the force acting on the particle.
The right-hand side is the collision operator.
The Boltzmann transport equation~\eqref{eq:transport} describes microscopic particle motion, while the Stokes equation~\eqref{eq:stokes} describes flow of a Newtonian continuum fluid.
However, both are based on the same conservation laws, those of mass and momentum,
and they can therefore be used interchangeably when studying mesoscopic hydrodynamics.

The LB method linearizes the collision operator $\tens{\Omega}$ on the right-hand side of \Cref{eq:transport} around the equilibrium distribution $f^\text{eq}$.
The total derivative on the left-hand side is discretized using finite differences\cite{duenweg09a} with a lattice spacing of  $\agrid$ and a time step of $\tgrid$ to obtain
\begin{equation}
	f_i(\vec{r}+\tgrid\vec{c}_i, t+\tgrid) = f_i(\vec{r},t)+\sum\limits_j \Omega_{ij} \underbrace{\left(f_j(\vec{r},t) - f^\text{eq}_j(\vec{r},t)\right)}_{f^\text{neq}_j(\vec{r},t)}
	\eqcomma
	\label{eq:lb}
\end{equation}
with the lattice vectors $\vec{c}_i$.

The collision represents a linear relaxation of the populations towards their equilibrium Maxwell-Boltzmann distribution.
At each lattice site, there is a number of fluid populations $f_i$, e.g., 19 different ones for the widely used D3Q19 model applied in this work, corresponding to a set of velocity vectors $\vec{c}_i$ connecting the site to neighboring cells.
For D3Q19, these are the rest mode, the six face neighbors, and the twelve edge neighbors.
In each time step, a collision operation is performed that redistributes fluid between the 19 populations within one cell, i.e., \cref{eq:lb} is applied with only $\tens{\Omega}$ on the right-hand side.
Next, a streaming step is performed which transfers the populations to the neighboring cell in the direction of their respective velocity vector, i.e., \cref{eq:lb} is applied without $\tens{\Omega}$ on the right-hand side.
From the populations $f_i$, the macroscopic physical quantities of fluid density and flow velocity are obtained using
\begin{align}
	\rho(\vec{r},t)&=\sum\limits_i f_i(\vec{r},t) \eqcomma \\
	\vec{u}(\vec{r},t)&=\frac{1}{\rho(\vec{r},t)}\sum\limits_i f_i(\vec{r},t)\vec{c}_i \eqdot
\end{align}

Throughout the manuscript, we choose a two relaxation time (TRT) LB\cite{ginzburg08a} collision operator for its high fidelity of reproducing no-slip (zero velocity) boundaries:
\begin{subequations}
	\begin{align}
		\sum\limits_j
		\Omega_{ij} f^\text{neq}_j(\vec{r},t)
		&= \lambda_\text{e} \left(f_i^+ - f_i^{\text{eq}+}\right) +
		\lambda_\text{o} \left(f_i^- - f_i^{\text{eq}-}\right) \eqcomma \\
		f_i^{\pm}(\vec{r},t) &= \frac{1}{2} \left( f_i(\vec{r},t) \pm f_{\bar{i}}(\vec{r},t) \right) \eqcomma \\
		f_i^{\text{eq}\pm}(\vec{r},t) &= \frac{1}{2} \left( f_i^\text{eq}(\vec{r},t) \pm f_{\bar{i}}^\text{eq}(\vec{r},t) \right)
		\eqdot
	\end{align}
\end{subequations}
$f_i^{\pm}$ and $f_i^{\text{eq}\pm}$ are symmetric ($+$) and antisymmetric ($-$) combinations of populations and equilibrium populations, respectively.
The symmetric relaxation parameter $\lambda_\text{e}$ derives from the fluid viscosity, while the antisymmetric $\lambda_\text{o}=3/16$ is tuned to minimize boundary slip\cite{ginzburg08b}.
The index $\bar{i}$ is the one whose velocity vector points in the direction opposite to the one with index $i$, i.e. $\vec{c}_{\bar{i}}=-\vec{c}_i$.

We will have forces acting on the fluid that vary in time and space for moving boundaries, therefore we need to augment the collision operation to be $\Omega_{ij}f^\text{neq}_j+\Delta_i$ with the appropriate force term $\Delta_i$.
We utilize the one originally constructed by \citet{guo02a} and generalized to multiple relaxation times\cite{schiller08a,schiller14a}:
\begin{subequations}
	\begin{align}
	\Delta_i &= \frac{3w_i\agrid^2\tgrid^2}{\rho} \left[\vec{f}_\text{ext}\cdot\vec{c}_i \phantom{\frac{\tgrid^2}{\agrid^2}}\right. \nonumber \\
	& \phantom{++}+ \left.\frac{3}{2} \mathrm{Tr}\left(
	\tens{G} \left( \vec{c}_i\otimes\vec{c}_i \right)\frac{\tgrid^2}{\agrid^2} - \frac{1}{3}\tens{G}
	\right)
	\right] \label{eq:guo} \eqcomma \\
	\tens{G} &= \frac{2+\lambda_\text{e}}{2} \left( \vec{u}\otimes\vec{f}_\text{ext} + \vec{f}_\text{ext}\otimes\vec{u} \right) 
	\eqcomma
	\end{align}
\end{subequations}
where $\otimes$ denotes the tensor (dyadic) product of two vectors and $\mathrm{Tr}$ the trace of a matrix.
$w_i$ is the lattice weight factor associated with direction $i$,
$\vec{f}_\text{ext}$ is the force acting on a lattice cell, and $\vec{u}$ the flow velocity in that cell.

Until we introduce the moving boundaries in \cref{sec:ladd}, we can consider colloidal particles by requiring them to be fixed relative to the lattice.
This corresponds to cells inside the colloid being a no-slip boundary for the fluid (corresponding to the co-moving frame, if a Galilean transformation is permitted).
In that case, the streaming step is modified so that populations streaming into the boundary are reflected back.
A generalization to arbitrary boundary velocities $\vec{v}_\text{b}$ is possible\cite{zou97a} and leads to
\begin{equation}
f_i(\vec{r}_\text{b}+\tgrid\vec{c}_i,t+\tgrid)
=f_{\bar{i}}(\vec{r}_\text{b},t)
+ \frac{6 w_i\tgrid^2}{\rho\agrid^2} \vec{c}_i\cdot \vec{v}_\text{b}
\eqcomma
\end{equation}
where $\vec{r}_\text{b}$ is the boundary node position.
This is referred to as a velocity bounce-back boundary.

We do not include thermal fluctuations of the solvent in this work.
Neither do we do so for the solute, as will be discussed in \cref{sec:ek}.
A consistent thermalization for the pure LB algorithm is available\cite{duenweg07a},
and such fluctuations may be significant for nanoscale systems.

\subsection{Moving LB Boundaries}
\label{sec:ladd}

There are multiple methods that incorporate particles into an LB fluid\cite{duenweg09a}, 
ranging from point-particle descriptions\cite{ahlrichs98a} to particles that are resolved on the grid\cite{ladd94a,aidun98a}.
In this paper, we are interested in the latter, for which 
the method by \citet{ladd94a} is the obvious choice to achieve coupling.
It considers all cells inside a particle as no-slip boundaries in the particle-co-moving frame.
In the lab frame, this corresponds to a velocity boundary condition with velocity $\vec{v}_\text{b}$ equal to the particle's surface velocity
\begin{equation}
	\vec{v}_\text{b}(\vec{r}_\text{b},t)=\vec{v}(t) + \vec{\omega}(t) \times (\vec{r}_\text{b} - \vec{r}(t))
	\eqdot
\end{equation}
Here, the particle is located at $\vec{r}$ and has linear velocity $\vec{v}$ and angular velocity $\vec{\omega}$. 
The linear and angular momentum exchanged during the bounce-back is transferred to the particle by applying an appropriate force and torque,
\begin{subequations}
	\label{eq:force}
	\begin{align}
	\vec{F}(t)&=\agrid^3\sum\limits_{\vec{r}_\text{b}} \sum\limits_{i=1}^{19} \vec{c}_i \left(f_i(\vec{r}_\text{b},t)+f_{\bar{i}}(\vec{r}_\text{f},t)\right) \eqcomma \\
	\vec{T}(t)&=\agrid^3\sum\limits_{\vec{r}_\text{b}} \sum\limits_{i=1}^{19} \left(\vec{r}_\text{b}-\vec{r}\right) \times \vec{c}_i \left(f_i(\vec{r}_\text{b},t)+f_{\bar{i}}(\vec{r}_\text{f},t)\right)
	\eqdot
	\end{align}
\end{subequations}
Here, the bounce-back is taking place between the boundary cell $\vec{r}_\text{b}$ and the fluid cell $\vec{r}_\text{f}$.
After determining these forces, the particle trajectory can be integrated, in our case using a symplectic Euler integrator.

These forces, along with any non-hydrodynamic forces applied, cause the particle to move across the lattice.
While it does so, the set of cells overlapped by the particle changes.
The original method by \citet{ladd94a} has some shortcomings, due to the presence of fluid inside the particle.
These include forces exerted on the particle by the internal fluid\cite{lowe95a} and the requirement that the particle's density is the same or larger than that of the fluid.
The modification by \citet{aidun98a} removes this unphysical fluid.
However, one now needs to destroy fluid in cells that are newly overlapped by a particle and create fluid in cells that are vacated.
The former is straightforward to do, while the latter requires choosing a value for each of the $f_i$ in the cell that becomes `new' fluid.

The initialization of the new fluid cell is problematic, as the D3Q19 LB model has more microscopic degrees of freedom than specified by the hydrodynamic boundary condition.
The typical choice is to use LB's equilibrium populations for a velocity equal to the particle's velocity and a density equal to the surrounding fluid cells' average density\cite{aidun98a}.
To conserve momentum, the created / destroyed fluid's momentum $\vec{p}_\text{f}$ is removed from / added to the particle as an additional force:
\begin{equation}
	\vec{F}(t)=\pm\frac{1}{\tgrid}\vec{p}_\text{f}(\vec{r}_\text{f},t)
	\eqdot
\end{equation}
Fluid mass is never conserved, as conserving it would violate the incompressibility of the fluid. However, the time-averaged fluid mass fluctuates around the correct value while the particle traverses the grid.

\subsection{Lattice Electrokinetics}
\label{sec:ek}

The method proposed by \citet{capuani04a} discretizes \cref{eq:flux} with a 19-point stencil on the same lattice that LB is performed on.
\citet{rempfer16a} have shown that the discretization originally proposed in Ref.~\onlinecite{capuani04a} introduces errors scaling exponentially with $E/\agrid$ where $E$ is the applied electric field.
We instead use
\begin{align}
	j_{ki}^\text{diff}&(\vec{r}\rightarrow\vec{r}+\vec{c}_i,t) \nonumber\\
	=& \frac{D_k}{\agrid}\left(\rho_k(\vec{r},t) - \rho_k(\vec{r}+\vec{c}_i,t)\right) \nonumber\\
	&- \frac{D_k z_ke}{k_\text{B} T\agrid(1+2\sqrt{2})} \frac{\rho_k(\vec{r},t) 
	+ \rho_k(\vec{r}+\vec{c}_i,t)}{2} \nonumber\\
	&\times \left(\Phi(\vec{r},t) - \Phi(\vec{r}+\vec{c}_i,t)\right)
	\label{eq:flux-disc}
\end{align}
for the diffusive term, as in Ref.~\onlinecite{rempfer16a}.
For the advective term $j_{ki}^\text{adv}$, a volume-of-fluid method is used\cite{capuani04a}: the solute in a cell at $\vec{r}$ is virtually displaced by a distance of $\vec{u}(\vec{r},t)\tgrid$, where $\vec{u}$ is the velocity of the LB fluid.
This displaced cell is intersected with all 26 neighboring cells to determine the amount of solute that needs to be transferred.

Once the fluxes have been calculated, they can be propagated using the discretized form of the continuity equation~\eqref{eq:continuity},
\begin{equation}
\frac{V_0}{\tgrid}\left(\rho_k(\vec{r},t+\tgrid)-\rho_k(\vec{r},t)\right) =
- A_0 \sum\limits_i j_{ki}(\vec{r},t)
\eqcomma
\end{equation}
with $V_0$ the cell volume and $A_0$ an effective surface area chosen such that \cref{eq:flux-disc} recovers the correct (bulk) diffusive mean-square displacement of $6Dt$. This results in $A_0=1+2\sqrt{2}$ for the D3Q19 lattice\cite{capuani04a}.

The fluid coupling force is a direct discretization of \cref{eq:fluidcoupling} and is applied using \cref{eq:guo}:
\begin{equation}
\vec{f}_\text{ext}(\vec{r},t)=\frac{k_\text{B}T\agrid}{\tgrid}\sum_k \sum_i \frac{j_{ki}^\text{diff}(\vec{r},t)}{D_k} \vec{c}_i
	\eqdot
\end{equation}

Electrostatics \eqref{eq:poisson} is treated using any available lattice-based electrostatics solver.
For this work, one based on Fast Fourier Transforms (FFT) is used, but iterative solvers for linear equation systems, such as Successive Over-Relaxation (SOR) or Krylov subspace methods, may also be applied.
The non-FFT-based solvers are also capable of incorporating inhomogeneous dielectric coefficients; the assumption of homogeneity has only entered into the specific form of Poisson's equation \eqref{eq:poisson} used here.
Besides driving the migrative flux \eqref{eq:flux-disc}, electrostatics also results in a force acting on the colloidal particles, which is added to the hydrodynamic drag force \eqref{eq:force} before integrating the particle trajectory.

At this point, EK is only capable of handling stationary boundary conditions, such as walls and particles that are stationary with respect to the lattice.
They are simply mapped into the LB fluid as a no-slip boundary in all cells inside them and any EK solute fluxes into or out of these cells is set to zero.
Using normal fluxes other than zero allows for the incorporation of chemical reactions occurring on the surface that lead to phenomena such as charge regulation and self-electrophoresis\cite{degraaf15c,brown15a}.
Boundaries may be charged by considering their charge distribution $\rho_\text{b}(\vec{r},t)$ as an additional summand when solving \cref{eq:poisson}.

As mentioned before, thermal fluctuations are not included.
The addition of noise to the concentration fields --- while obtaining the proper fluctuation-dissipation relation for the total fluid --- is non-trivial and goes beyond the scope of the current manuscript.
Thermalization of the EK algorithm based on existing numerical methods for fluctuating hydrodynamics of electrolytes\cite{peraud16a,zudrop14a} shall be the subject of future study.

\subsection{Moving EK Boundaries}
\label{sec:ek-moving}

In \cref{sec:ladd}, the well-known moving boundary method for LB was introduced.
It is not mass-conserving, so a straightforward adaptation to EK would cause the amount of solute to vary over time and thus violate charge conservation.
Furthermore, as the solute charge is typically far less homogeneously distributed than the fluid mass, one can expect that charge would not even be conserved on average over long time scales.
A simulation undergoing such a net charge drift will typically not be able to produce physically correct results.

\subsubsection{Simple Charge Conservation Scheme}
\label{sec:simplescheme}

\begin{figure}[b]
	\centering
	\includegraphics[width=\linewidth]{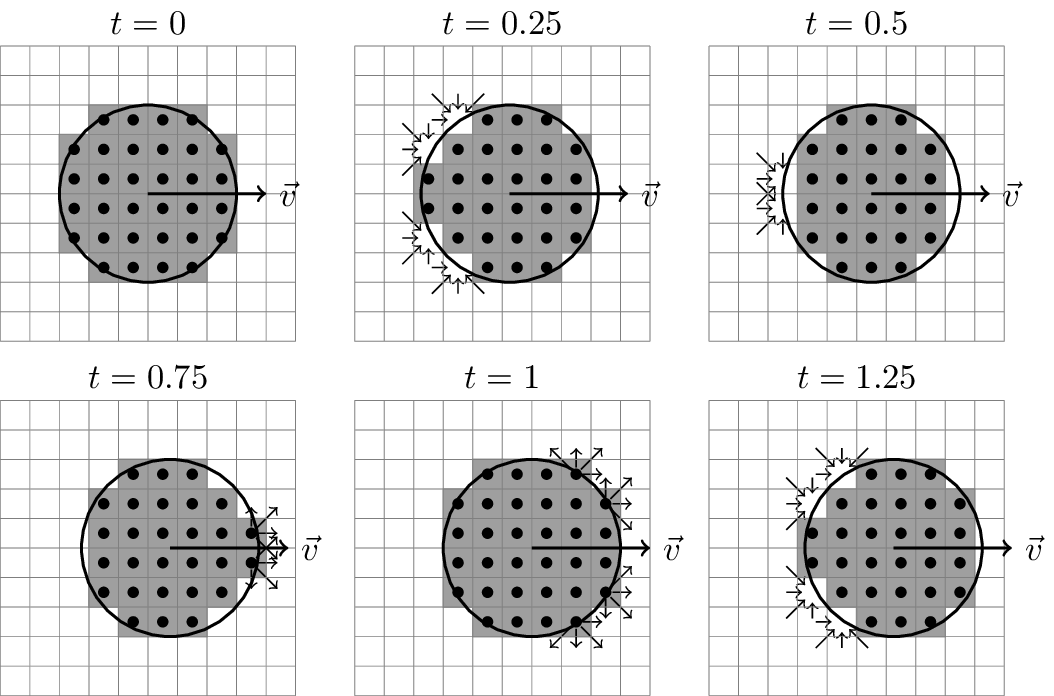}
	\caption{Illustration of the mass conservation modification to the Ladd boundary scheme to make it usable for EK. Cells whose center is inside the particle are considered to be boundary nodes. The arrows indicate how solute is drawn into vacated cells (panes 2, 3, and 6) and expelled from newly-overlapped cells (panes 4 and 5).}
	\label{fig:chargecons}
\end{figure}

To avoid a net charge drift, any solute from cells claimed by a particle needs to be expelled and redistributed into surrounding cells.
Such a scheme is illustrated in \cref{fig:chargecons}.
Cells vacated behind the particle are refilled by taking the sum of the solute concentration of the surrounding $N_\text{f}$ non-boundary cells and dividing it by $N_\text{f}+1$ to account for the fact that this amount of solute will now be shared with one additional cell.
This means that the solute concentration
\begin{equation}
	\rho_k(\vec{r}_\text{new},t+\tgrid)=\frac{1}{N_\text{f}+1}\sum\limits_{i=1}^{N_\text{f}} \rho_k(\vec{r}_\text{new}+\vec{c}_i,t)
	\label{eq:refill}
\end{equation}
is put into the vacated cell.
To conserve total solute mass, that amount is then removed from the surrounding non-fluid cells in amounts proportional to their current solute concentration:
\begin{equation}
	\rho_k(\vec{r}+\vec{c}_i,t+\tgrid) = \rho_k(\vec{r}+\vec{c}_i,t) -\frac{\rho_k(\vec{r}+\vec{c}_i,t)}{N_\text{f}+1}
	\label{eq:refillneighbors}
\eqdot
\end{equation}
It should be noted that expulsion and vacation do not necessarily happen at the same time in the front and back of a moving particle, and thus the total number of boundary cells oscillates over time.

\subsubsection{Enhanced Partial Volume Scheme}

\begin{figure}[b]
	\centering
	\includegraphics[width=\linewidth]{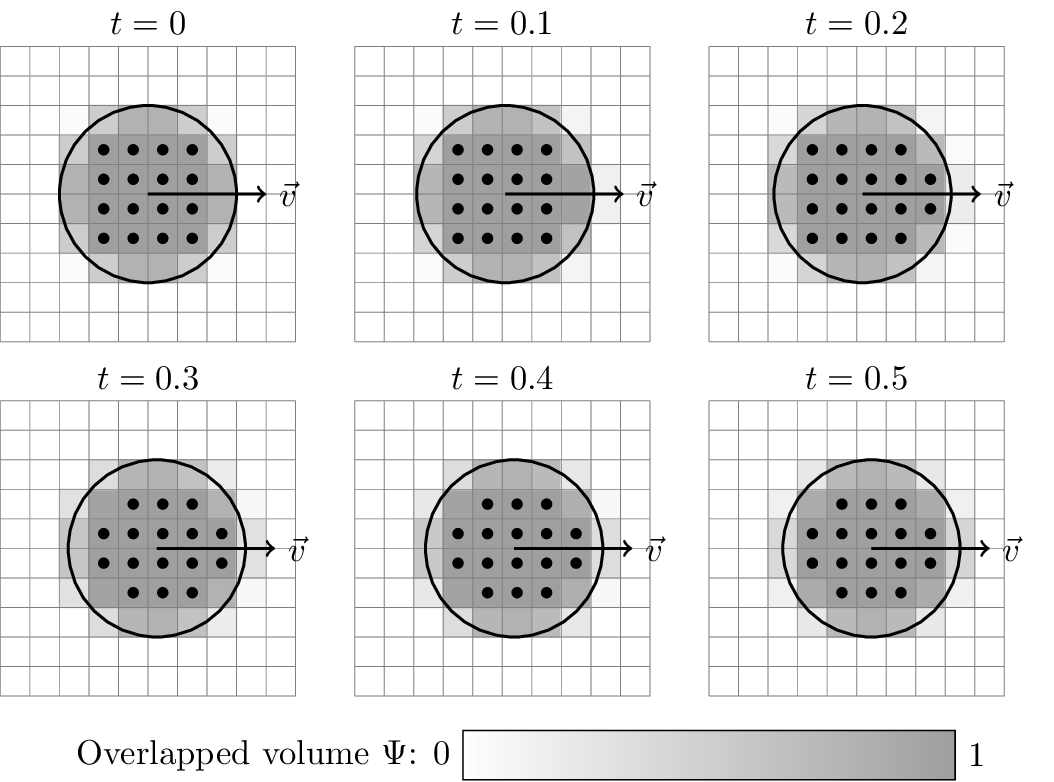}
	\caption{Illustration of the partial volume scheme for moving boundaries in EK. The shading of the cells inside the particle corresponds to the overlapped volume $\Psi$ to indicate how the particle's charge is distributed across the cell layer at its surface. In the calculation of the diffusive flux, the concentrations are scaled with $1-\Psi$ to determine the effective concentrations.}
	\label{fig:partialvolume}
\end{figure}

The scheme introduced in \cref{sec:simplescheme} has the disadvantage that it moves large amounts of solute at a few points in time.
Therefore, as we demonstrate in \cref{sec:validation}, the particle's velocity can vary quite strongly during the time steps after a cell has been claimed or vacated.
To reduce these effects, we propose a partial volume scheme, which is illustrated in \cref{fig:partialvolume}.

In the following, $\Psi(\vec{r},t)$ is a field describing the volume fraction of the cell at $\vec{r}$ that is overlapped by a particle, with $\Psi=1$ meaning that the cell is completely inside the particle and $\Psi=0$ completely outside.
In the calculation of the diffusive fluxes \eqref{eq:flux-disc}, the concentrations are replaced with ones that take into account that all solute resides in the non-overlapped part of the cells.
To prevent the resulting diffusive fluxes from diverging as $\Psi\rightarrow 1$, we renormalize them by scaling them with the volume.
This leads to the following modified expression for the flux:
\begin{align}
	j_{ki}^\text{diff}&(\vec{r}\rightarrow\vec{r}+\vec{c}_i,t) \nonumber\\
	=& \left[ \frac{D_k}{\agrid}\left(\frac{\rho_k(\vec{r},t)}{1-\Psi(\vec{r},t)} - \frac{\rho_k(\vec{r}+\vec{c}_i,t)}{1-\Psi(\vec{r}+\vec{c}_i,t)}\right) \right. \nonumber\\
	&- \frac{D_k z_ke}{2 k_\text{B} T\agrid(1+2\sqrt{2})} \left( \frac{\rho_k(\vec{r},t)}{1-\Psi(\vec{r},t)} \right. \nonumber\\
	&+ \left.\frac{\rho_k(\vec{r}+\vec{c}_i,t)}{1-\Psi(\vec{r}+\vec{c}_i,t)} \right)
	\times \left. \vphantom{\frac{1}{1}} \left(\Phi(\vec{r},t) - \Phi(\vec{r}+\vec{c}_i,t)\right) \right] \nonumber\\
	& \times \left(1-\Psi(\vec{r},t)\right) \left(1-\Psi(\vec{r}+\vec{c}_i,t)\right)
	\label{eq:flux-pv}
	\eqdot
\end{align}
With this change, refilling vacated cells as per \cref{eq:refill,eq:refillneighbors} is no longer necessary. They can be set to zero concentration and will be filled up by the diffusive flux again as $\Psi$ increases.
We calculate $\Psi$ numerically by sub-dividing each cell into 8 equally-sized cells and determining how many of them are completely inside and completely outside the particle.
For those cells that are neither, the subdivision is recursively repeated up to a maximum depth of $4$.
Expelling solute from a cell that is claimed by a particle is, however, still necessary --- even with the modified expression for the flux --- as the cell is not necessarily completely empty by the time it is claimed, due to the discretized motion of the colloid.
The expelled amount of solute with \cref{eq:flux-pv} is much smaller than with \cref{eq:flux-disc} and thus the effect of this sudden change on the simulation is reduced to acceptable levels.

One further source of sudden variations in solute fluxes is the change in electrostatic potential when the volume across which a particle's charge is distributed varies due to the fluctuation in the number of boundary cells.
Therefore, when calculating the electrostatic potential, each particle's total charge $Q=Ze$ is distributed among all cells that are at least partially overlapped by that particle:
\begin{equation}
	\rho_\text{b}(\vec{r},t) = Ze\frac{\Psi(\vec{r},t)}{V_\text{p}}
	\eqcomma
\end{equation}
with $V_\text{p}$ the particle's (non-discrete) volume.
Inhomogeneous charge distributions are also possible as long as the charge in a cell varies smoothly while the cell is slowly claimed or vacated by the colloidal particle.

\section{Validation}
\label{sec:validation}

We implement our new algorithm using the \textsf{waLBerla} framework\cite{godenschwager13a}.
It supports several lattice Boltzmann models, including the one introduced in \cref{sec:lb}, and correctly handles the moving LB boundaries described in \cref{sec:ladd}.
We already added an implementation of the EK model described in \cref{sec:ek}.
\textsf{waLBerla} provides excellent scaling on high-performance computing clusters and contains advanced features, such as grid refinement\cite{schornbaum16a}, that may prove useful for EK simulations of complex systems in the future.

\subsection{Electrophoresis of a Single Colloid}
\label{sec:singlecolloid}

To validate the moving boundary EK method described in \cref{sec:ek-moving}, we choose a simple system consisting of an electrophoresing sphere.
Specifically, we simulate a single, homogeneously charged, spherical colloid in a cubic box with periodic boundary conditions (PBCs) undergoing electrophoresis in a uniform external electric field.
This system has already been studied extensively using the EK method without moving boundaries by considering the equivalent problem of electro-osmotic flow\cite{giupponi11a}.

The parameters used are given in \cref{tab:parameters}.
They were chosen to match one of the data points of Ref.~\onlinecite{giupponi11a}.
A salt concentration of $c=\SI{1e-3}{\mole\per\litre}$ is experimentally relevant\cite{lobaskin07a,garbov04a,medebach04b,palberg04a} and the (effective) colloid charge of $Z=30$ is realistic for a particle of radius $R=\SI{4e-9}{\metre}$.
The resulting Debye length of $\lambda_\text{D}=\SI{10}{\nano\metre}$ for that concentration is neither in the thin ($\lambda_\text{D}\ll R$) nor thick ($\lambda_\text{D}\gg R$) double layer limit and should thus demonstrate the capability of the method to deal with these intermediate regimes.
The charge of $Z=30$ was chosen such that Debye-H\"uckel theory is not applicable --- the requirement being $e\zeta < k_\text{B}T$ with $\zeta=Zk_BT\lambda_\text{B}/(R+R^2/\lambda_\text{D})$ --- while for $Z=3$, it is.
\begin{table}
\centering
\begin{tabular}{c|ll}
Temperature & $T=\SI{298.15}{\kelvin}$ & \\
Relative permittivity & $\varepsilon=78.54$ &  \\
Fluid density & $\rho=\SI{997.04}{\kilo\gram\per\metre\cubed}$ & \\
Viscosity & $\eta=\SI{0.8937e-3}{\pascal\second}$ &  \\
\multicolumn{3}{c}{} \\
Salt concentration 
                   & $c=\SI{1e-3}{\mole\per\litre}$ & \\
                   & $c=\SI{1e-4}{\mole\per\litre}$ & \\
Salt valency & $z_\pm=\pm 1$ & \\
Diffusion coefficient & $D=\SI{2}{\metre\squared\per\second}$ & \\
External field & $E=\SI{256.9e3}{\volt\per\metre}$ & \\
\multicolumn{3}{c}{} \\
Sphere radius & $R\approx\SIrange{3e-9}{8e-9}{\metre}$ & \\
Sphere charge & $Q=\SI{3}{\elementarycharge}=\SI{0.4807e-18}{\coulomb}$ & \\
              & $Q=\SI{30}{\elementarycharge}=\SI{4.8807e-18}{\coulomb}$ & \\
Density of particle & $\rho_\text{p}=2\rho$ & \\
Box length & $L=\SI{64e-9}{\metre}$ & \\
\multicolumn{3}{c}{} \\
Length unit & $\Delta x = \SI{1e-9}{\metre}$ & \\
Mass unit & $\Delta m = \rho \Delta x^3=\SI{9.97e-25}{\kilo\gram}$ & \\
Energy unit & $\Delta E=k_\text{B}T=\SI{4.12e-21}{\joule}$ & \\
Time unit & $\Delta t=\sqrt{{\Delta m\Delta x^2}/{\Delta E}}$ & \\
          & $\phantom{\Delta t}=\SI{1.56e-11}{\second}$ & \\
LB grid spacing & $\agrid=\Delta x = \SI{1e-9}{\metre}$ & \\
LB time step & $\tgrid=0.2\Delta t=\SI{3.1e-12}{\second}$ & \\
\end{tabular}
\caption{Parameters used in the simulation and their conversion to simulation units.
Note that our choice of $\rho_\text{p}=2\rho$ is arbitrary and irrelevant to the working of our algorithm, as verified for a single system where we used $\rho_\text{p}=\rho$.}
\label{tab:parameters}
\end{table}
The salt concentration $c$ and the grid spacing $\agrid$ need to be chosen such that one Debye length is resolved by a minimum of approximately 4 cells, as we will see during validation.
This precludes us from using a salt concentration of much higher than $c=\SI{1e-3}{\mole\per\litre}$ without increasing the grid resolution.
However, sufficiently resolving the double layer is a general requirement of the EK method\cite{capuani04a}, as well as for FEM and other algorithms, and not specific to the use of moving boundaries.
We will compare the results of the different simulation methods via a single number, the reduced electrophoretic mobility\cite{wiersema66a}:
\begin{equation}
	\tilde\mu=\frac{6\pi \eta \lambda_\text{B}}{e}\frac{v}{E}
	\eqcomma
\end{equation}
where $v$ is the speed of the particle's motion relative to the bulk fluid and $E$ is the applied electric field.

\begin{figure}[t]
	\centering
	\includegraphics[width=\linewidth]{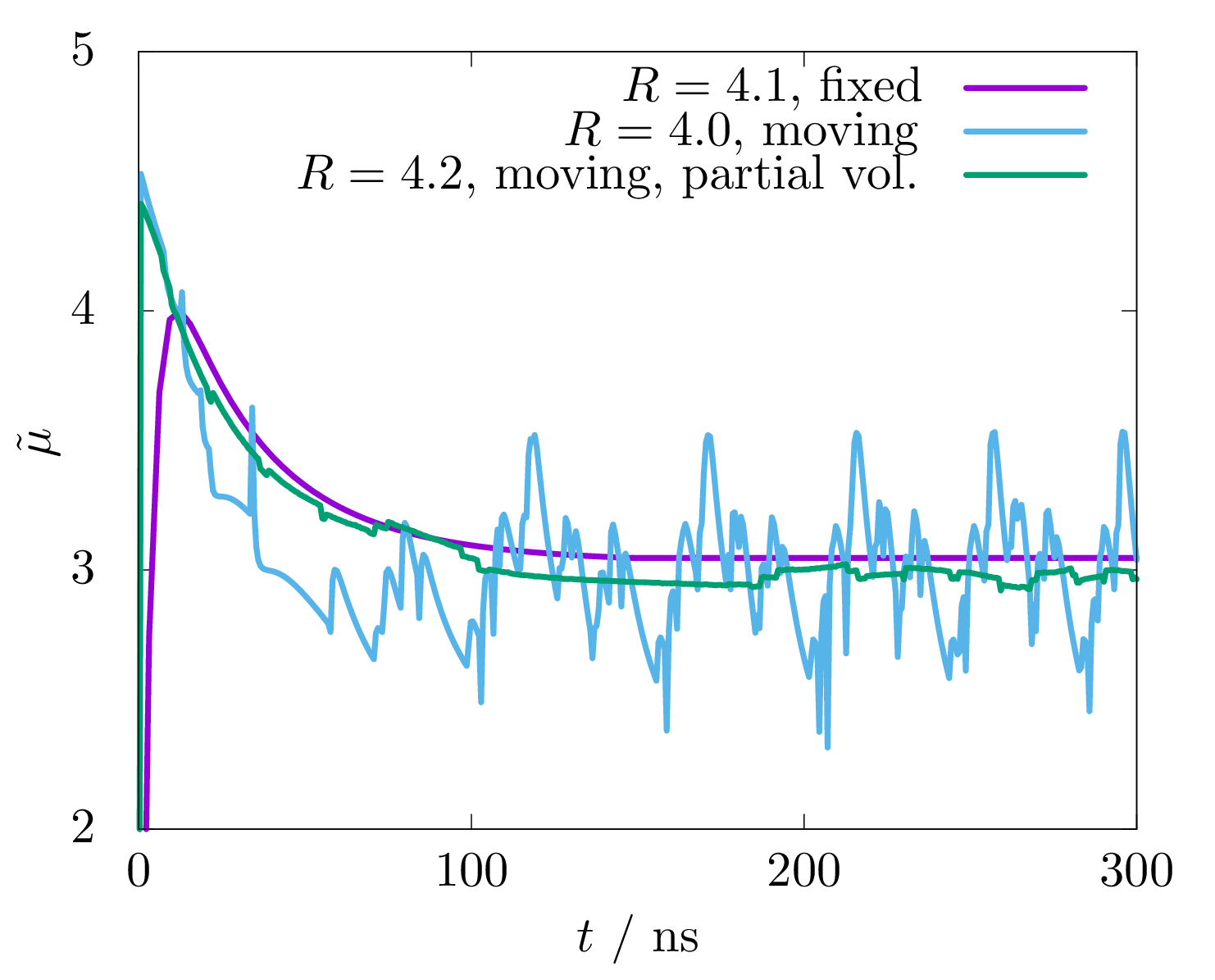}
	\caption{Comparison of the transient behavior ($t$ denotes time) of the reduced electrophoretic mobility $\tilde\mu$ for the fixed particle (purple curve), the particle moving via the simple moving boundary scheme (blue curve), and the particle moving via the partial volume scheme (green curve).
	Here, $R=\SI{4e-9}{\metre}$, $Z=30$, and $c=\SI{1e-3}{\mole\per\litre}$ were used.}
	\label{fig:bothmethods}
\end{figure}

\Cref{fig:bothmethods} compares results obtained for the same parameter set (particle radius $R\approx\SI{4e-9}{\metre}$, particle charge $Z=30$, and salt concentration $c=\SI{1e-3}{\mole\per\litre}$) for a simulation with a fixed particle, a moving particle without, and a moving particle with the partial volume scheme, respectively.
The effective particle radius in the moving scheme is obtained by counting the number of cells that are completely inside the particle, averaging that number over the simulation time, and setting it equal to the sphere volume $(4/3)\pi R^3$.
This allows us to choose $R$ such that the moving and stationary simulations study the same physical system.
One can see that both moving particle methods on average agree with the fixed boundary simulation's stationary mobility value to within $1.4\%$ and $2.4\%$, respectively.
Agreement of the full time evolution is not expected as the electro-osmotic (fixed boundary) and electrophoretic (moving boundary) problems are, in fact, only equivalent in the stationary state.
The method without partial volumes results in a mobility oscillating with an amplitude of $20.3\%$ around the mean and a period corresponding to the time it takes the particle to move forward by one grid cell.
The partial volume scheme on the other hand reduces the mobility oscillations to $1.7\%$, which is generally small enough to consider any instantaneous value to be a good estimate for the true value.
The period remains bound to the particle moving forward by one cell.

\begin{figure}[b]
	\centering
	\includegraphics[width=\linewidth]{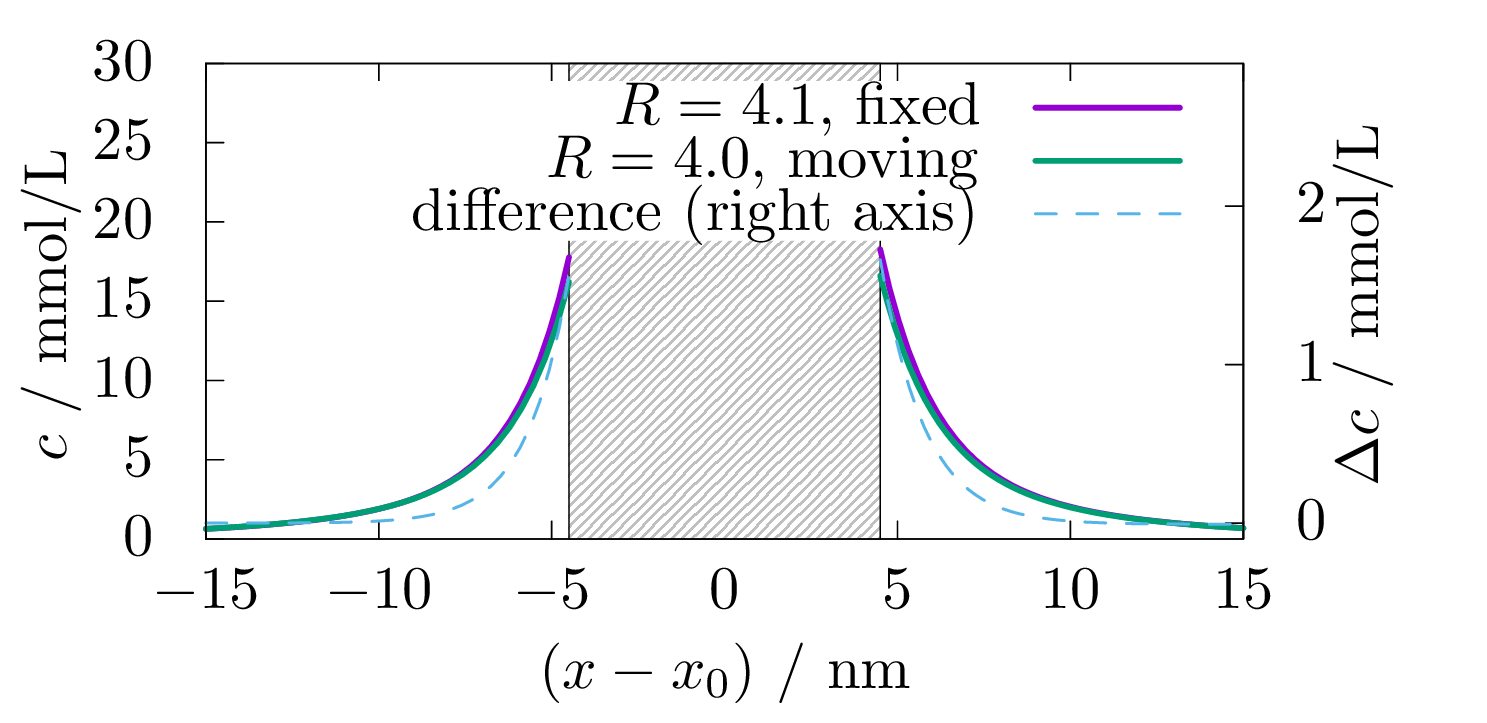}
	\caption{Comparison of the shape of the double layer around the colloid for the fixed and moving particles.
	Here, $R\approx\SI{4e-9}{\metre}$, $Z=30$, and $c=\SI{1e-3}{\mole\per\litre}$ were used.
	The shaded area refers to the space taken up by the colloid.
	The dashed line shows the absolute difference between the two charge distributions.}
	\label{fig:chargedist}
\end{figure}

To ensure that the method does not adversely affect the shape of the electric double layer around the colloid, we examine the charge distribution around the colloid in \cref{fig:chargedist}.
We see there is a small deviation close to the colloid's surface, but we can attribute this to the slight difference in particle size, as discussed above.

\begin{figure}[t]
	\centering
	\includegraphics[width=\linewidth]{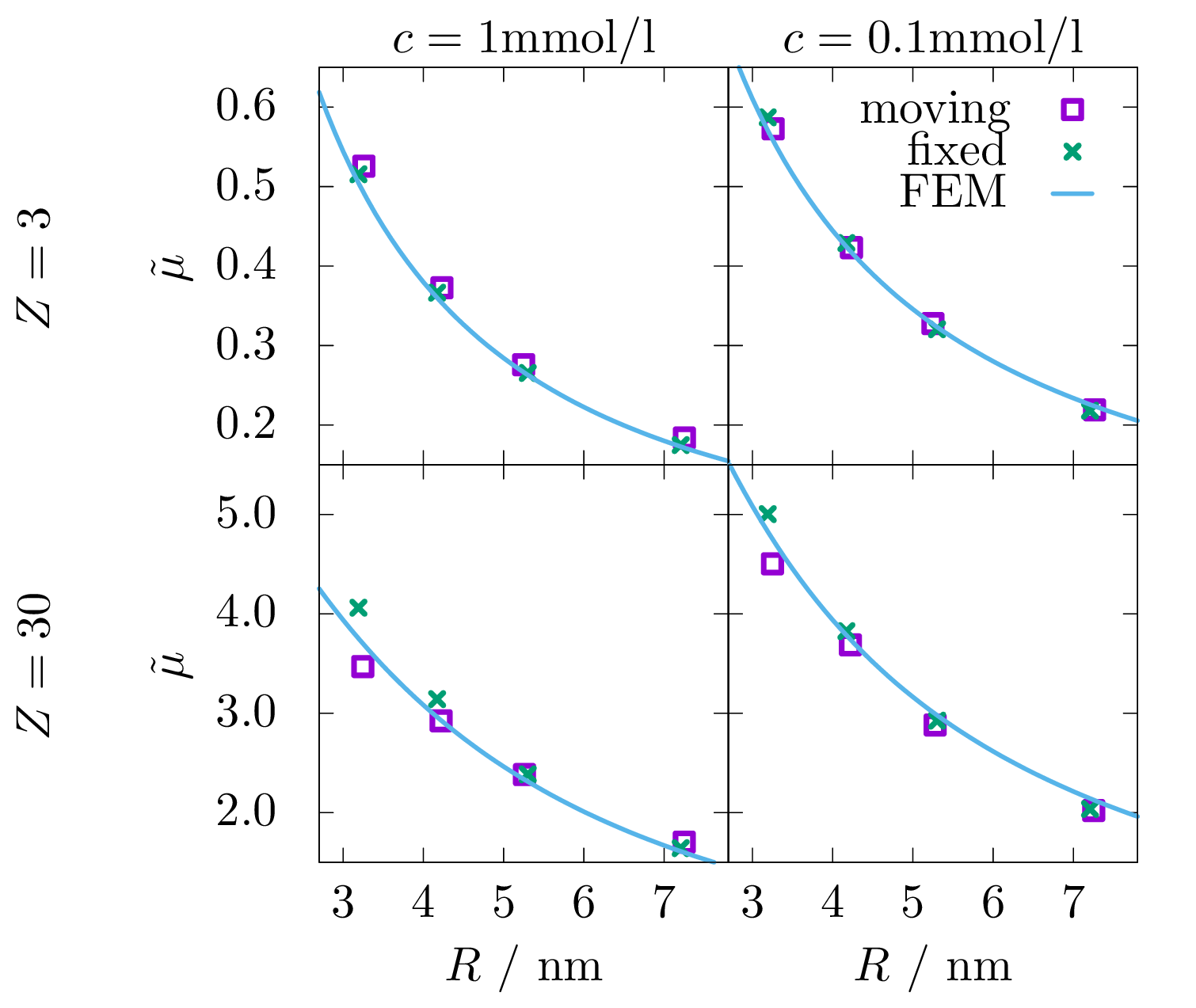}
	\caption{Comparison of the steady-state value of the electrophoretic mobility for different values of $R$, $c$, and $Z$.}
	\label{fig:results}
\end{figure}

We further validated the partial volume method by running additional simulations with various concentrations $c$, particle charges $Z$, and particle radii $R$.
The results are presented in \cref{fig:results} and are compared with reference results obtained by solving the electrokinetic equations with the FEM solver \textsf{COMSOL}.
Note that we use a finite but large simulation domain in the FEM model, while the EK model handles PBCs naturally; this may lead to slight differences in the results obtained with both methods.
We observe an excellent agreement between fixed and moving boundary simulations, as well as the FEM reference results.
The agreement improves as the particle size is increased, and thus the quality of the staircase approximation to its spherical shape.
For example, for $R\approx\SI{7e-9}{m}$, $Z=30$, and $c=\SI{1e-3}{\mole\per\litre}$, the oscillation amplitude drops to $1.3\%$.
The remaining difference in the mobility is easily explained by a slight mismatch in volume (and thus effective radius) between the fixed and the moving particle.

Finally, we examined the speed of our simulation compared to the method of \citet{capuani04a} for the geometries that we considered.
The moving boundary algorithm adds only 2\% to the simulation time per time step, when compared to the equivalent fixed boundary simulation.
Our smoothing results in a further increase of less than 10\% in time, when compared to the non-smoothed moving boundaries.
Together, this shows that our method does not incur an unreasonable computational cost and can therefore be straightforwardly applied to domain sizes that are currently accessible to the \citeauthor{capuani04a} method\cite{capuani04a,giupponi11a}. 

\subsection{Electrophoresis of a Colloidal Suspension}

\begin{figure}[t]
	\centering
	\includegraphics[width=\linewidth]{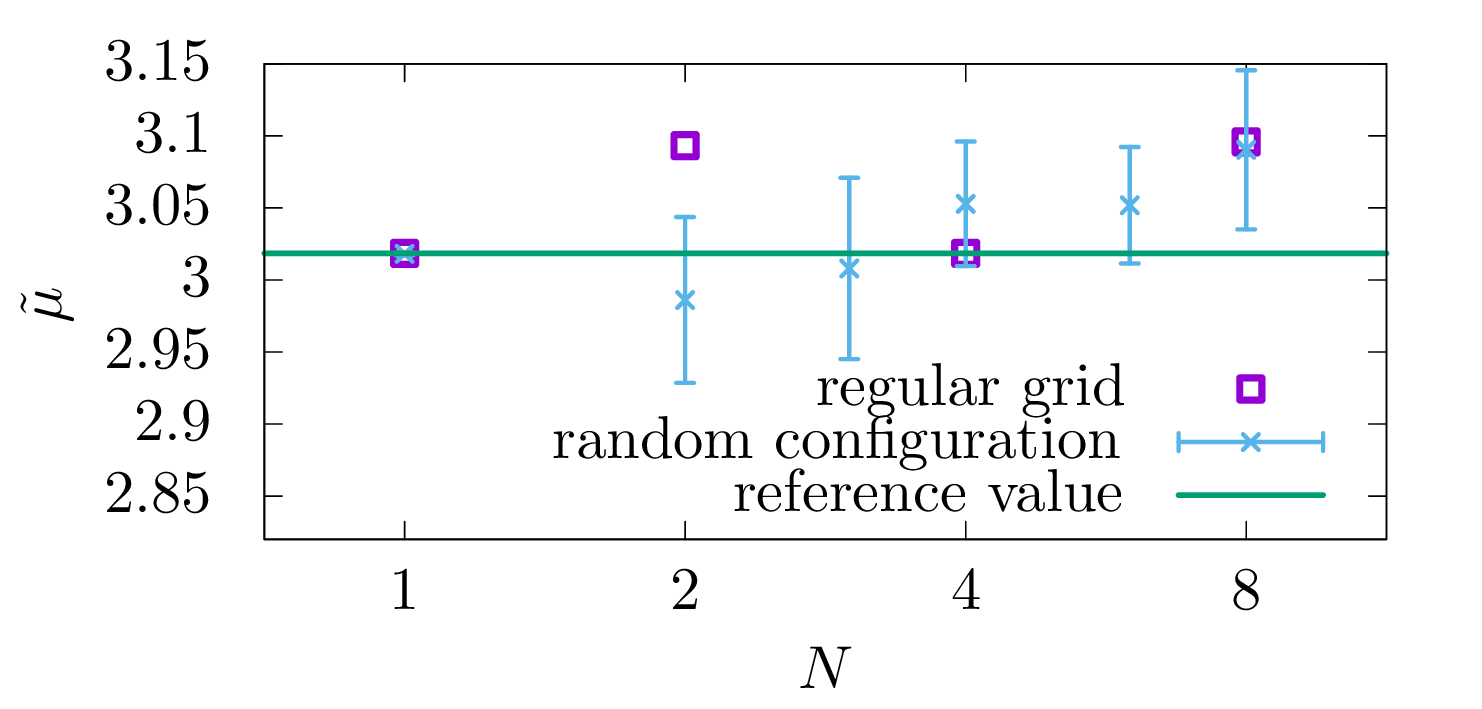}
	\caption{Comparison of the steady-state value of the electrophoretic mobility for different numbers of colloids at constant volume fraction.
	Here, $R\approx\SI{4e-9}{\metre}$, $Z=30$, and $c=\SI{1e-3}{\mole\per\litre}$ were used.
	The error bars show the spread of values for different initial random placements of colloids in cubic boxes, while the squares show the value obtained for initial placement on a regular lattice.}
	\label{fig:multiple}
\end{figure}

In this section, we consider the electrophoresis of a system comprised of several spheres.
This allows us to verify our moving boundary method in a more physically complex system with interactions that the original \citeauthor{capuani04a} method could not access.
We pick one parameter set ($R\approx\SI{4e-9}{\metre}$, $Z=30$, and $c=\SI{1e-3}{\mole\per\litre}$) and repeat the simulation from \cref{sec:singlecolloid}.
At these parameters, the volume fraction is about 0.1\% and the double layers do not overlap noticeably, therefore only long-ranged hydrodynamic interactions are expected to mediate interactions between the colloids.
Instead of having one spherical colloid interact with only its periodic images, we now enlarge the simulation box and add additional identical spheres while keeping the colloidal volume fraction constant.
The colloids are either positioned randomly in a cubic box or in a regular lattice in a cuboidal box.
This kind of simulation is typically performed to ensure the results do not suffer from artifacts of the PBCs\cite{lobaskin07a}.
As opposed to the single colloid and its periodic images, which always maintain their relative positions, the colloids in this simulation are free to move relative to each other.
\Cref{fig:multiple} shows that the electrophoretic mobility is indeed mostly independent of the number of colloids, varying $0.9\%$ from the average and $1.8\%$ from the value for one colloid in PBCs.
This matches the findings of \citet{lobaskin07a} using colloids modeled with the Ahlrichs-D\"unweg coupling\cite{ahlrichs98a,fischer15a,degraaf15b}, where agreement to within a few percent was found.
The results of \cref{sec:singlecolloid} are therefore not artifacts of the periodicity of the simulation domain.
For the case of a cubic simulation domain, we furthermore observe that the alignment of the colloids on a regular body-centered cubic lattice is a stable configuration.

\section{Conclusion}
\label{sec:conclusion}

Summarizing, we have introduced a method to simulate electrokinetic phenomena in colloidal suspensions.
Our scheme builds upon the lattice electrokinetics algorithm of \citet{capuani04a},
which itself is capable of simulating moving colloids only by considering them as stationary boundary conditions in the Galilean-transformed (co-moving) frame.
Motion of the colloids relative to the lattice can be incorporated by employing a method similar to the moving boundary method for LB\cite{ladd94a}.
A key modification is required, however: We introduce mass conservation for the solute species in order to conserve charge.
This is accomplished by redistributing solute from and to cells neighboring the ones that were recently vacated and claimed by the particle, respectively.

The above procedure is, in principle, sufficient to enable the simulation of, for example, colloidal electrophoresis.
However, in practice, further improvements are desirable to allow for the simulation of colloids with as few LB cells as possible.
To reduce the effects of sudden and strong solute fluxes, when a cell is claimed or vacated by a particle, a smoothing scheme by partial volumes is introduced. That is, the electrokinetic equations are solved for an effective concentration that incorporates what fraction of a cell is actually accessible to the solute, i.e., not overlapped by a colloid.
This form of smoothing reduces the lattice artifacts by more than a decade in our test simulations.
We also showed that having smoothing and moving boundaries only slows down the simulation by about $10\%$ compared to the original Capuani algorithm.

We validated our method for the electrophoresis of spherical colloids.
For a single sphere, we find excellent agreement of the reduced steady-state electrophoretic mobility $\tilde\mu$ with the one obtained for the equivalent problem of electro-osmotic flow around a fixed sphere (co-moving frame).
One can also obtain a reliable estimate for the mobility without the smoothing via partial volumes, but in this case it is necessary to average $\tilde\mu$ over one period of its oscillations.
While this is a feasible solution for the system under consideration here, systems where the transient behavior of multiple particles is of interest would require averaging over many periods, thus requiring much longer overall simulation times.
Alternatively, the oscillations could be reduced by increasing the grid resolution.
This is again not desirable, as it comes with a steep increase in required computation time --- performing the same simulation at twice the grid resolution takes eight times as much computing time.
It is therefore clear that our smoothing is a prerequisite for the efficient study of electrokinetic moving boundary problems.

Our moving boundary EK method was further verified for the electrophoresis of multiple spheres that were free to move and interact.
We find the electrophoretic mobility of the spheres to be almost independent of their initial arrangement.
Such a simulation can only be performed with moving boundaries and thus demonstrates the power of the method presented.

Note that our moving boundary EK method imposes no limitations on simulation parameters beyond those already present in the original non-moving algorithm.
These requirements include sufficient discretization of the double layer and limitations on the choice of time step, diffusivity, and grid discretization\cite{capuani04a}.
Furthermore, it leaves the EK and LB algorithms almost unmodified, which makes it easy to incorporate into existing LB or EK simulation codes.
Finally, it also does not influence the scaling behavior of these algorithms as the work required to map a particle onto the lattice is linear in the number of lattice cells and particles.

In this investigation, we have primarily studied the case of external electrophoresis of a single colloid.
As we have shown, however, our method can also be employed to investigate cases where Galilean transformation to a co-moving frame for the study of an equivalent electro-osmotic is not possible.
These types of systems will be the subject of future work and could include a diverse range of systems.
For example, our method is readily applicable to study the translocation, characterization, concentration, separation, and transport of DNA, proteins, and other biochemical analytes; interactions between oppositely charged nanoparticles in an oscillatory electric field; collective dynamics of self-electrophoresing colloids, and many others.
Our work thus opens the door for the study of a wide range of physical systems that were previously inaccessible to continuum lattice-based methods.

\begin{acknowledgments}
MK, GR, JdG, and CH thank the DFG for funding through the SPP 1726  ``Microswimmers --- From Single Particle Motion to Collective Behavior''.
GR and CH acknowledge further funding through the SFB716, TP C.5.
JdG gratefully acknowledges financial support by an NWO Rubicon Grant (\#680501210)
and funding by a Marie Sk\l odowska-Curie Intra European Fellowship (G.A. No. 654916) within Horizon 2020.
We thank G. Davies and J. Harting for useful discussions.
\end{acknowledgments}

\bibliography{kuron16a,bibtex/icp,georg}

\end{document}